\documentclass[12pt,preprint]{aastex}
\usepackage{epstopdf}
\usepackage{amsmath}
\usepackage{hyperref}
\usepackage{float}
\usepackage{color}

\slugcomment{PASP, September 2013 edition}
\shorttitle{Bad pixel interpolation}
\shortauthors{Popowicz et al.}

\begin{document}

\title{Bad pixel modified interpolation for astronomical images}

\author{A. Popowicz\altaffilmark{1}}
\affil{Silesian University of Technology, Institute of Electronics, Gliwice, Poland}
\email{apopowicz@polsl.pl}

\author{A. R. Kurek\altaffilmark{2}}
\affil{Astronomical Observatory of the Jagiellonian University, Krakow, Poland}
\email{aleksander.kurek@uj.edu.pl}

\author{Z. Filus}
\affil{Silesian University of Technology, Institute of Electronics, Gliwice, Poland}
\email{zfilus@polsl.pl}

\altaffiltext{1}{Main author.}
\altaffiltext{2}{Second author.}

\begin{abstract}

We present a new method of interpolation for the pixel brightness estimation in astronomical images. Our new method is simple and easily implementable. We show the comparison of this method with the widely used linear interpolation and other interpolation algorithms using one thousand astronomical images obtained from the Sloan Digital Sky Survey. The comparison shows that our method improves bad pixels brightness estimation with four times lower mean error than the presently most popular linear interpolation and has a better performance than any other examined method. The presented idea is flexible and can be also applied to presently used and future interpolation methods. The proposed method is especially useful for large sky surveys image reduction but can be also applied to single image correction.

\end{abstract}


\section{Introduction}

Astronomical images taken with image sensors are nowadays one of the most important tools of the modern astronomy \citep{saha09, mclen08}. The most popular are CCD and CMOS sensors which consist of a matrix of pixels, where the light flux can be measured thanks to the photovoltaic effect \citep{lit01, jan01, hob78}. 
Unfortunately, not every pixel can be used effectively. It is due to the possible imperfections located within the pixel. The most commonly encountered problems are: high dark current, rate saturating the pixel's potential well; nonlinear dark current dependencies \citep{wid10, dun12, pop11a, pop11b}; transient events of the dark current due to the irradiation (especially important for flight missions) \citep{hop08, hop96}, pixel nonlinear light response and CCD fabrication defects \citep{jan01}. In professional CCD systems there are several methods developed for investigation of the so-called bad pixels \citep{hil12a, hil12b}. The bad pixel masks are created to help during the image reduction. This calibration is usually repeated periodically because new bad pixels can appear. 
One way to reduce bad pixels impact on the image quality is to take many images with slight shifts. Such a set of dithered pictures is shifted back and averaged, ignoring the pixels from the mask \citep{fru02}. Unfortunately, it is not possible if there is only a single image and in such cases an interpolation over bad pixels is necessary. The most popular method is a linear interpolation which is a standard procedure in the most widespread astronomical software package: {\it Image Reduction and Analysis Facility} (IRAF) \citep{mas97}. 
In this paper, we propose a new method of bad pixel interpolation for astronomical images. In chapter ~\ref{sec:2} we present the data which was chosen for the methods comparison. Next, in chapter ~\ref{sec:3.1}, we describe most popular interpolation methods and we also introduce biharmonic interpolation algorithm as a new alternative for bad pixel correction. In chapter ~\ref{sec:3.2} we summarize the methodology of our tests and discuss the results. In chapter ~\ref{sec:4.1} we present our new method which is a modification of the biharmonic interpolation. The modification is dedicated especially for astronomical pictures calibration. We suggest some further enhancements which improve its effectiveness and precision. The result of our modified method compared to the original biharmonic interpolation are presented in chapter ~\ref{sec:4.2}. We conclude in section ~\ref{sec:5}.

\section{Data}
\label{sec:2}

The verification data set made use of the astronomical images obtained by the Sloan Digital Sky Survey (SDSS) Data Release 7 \citep{yor00, szal99}. The SDSS uses 2.5m Ritchey-Chretien telescope located at Apache Point Observatory in New Mexico. The imaging camera consists of an array of 30 CCD with total field of view of 3 degrees and operating in drift-scan imaging mode.
We used {\it SDSS Science Archive Server} \footnote{\url{http://dr9.sdss3.org/}} (SAS) to download 1000 astronomical images. The images were previously calibrated by standard SDSS pipeline reduction procedures. We chose SDSS $g$ filter images located at random positions on the sky (randomized RA and DEC coordinates). We present exemplary image in Fig. 1.

\begin{figure}[H]
\label{fig:1}
\centering
\epsscale{0.65}
\plotone{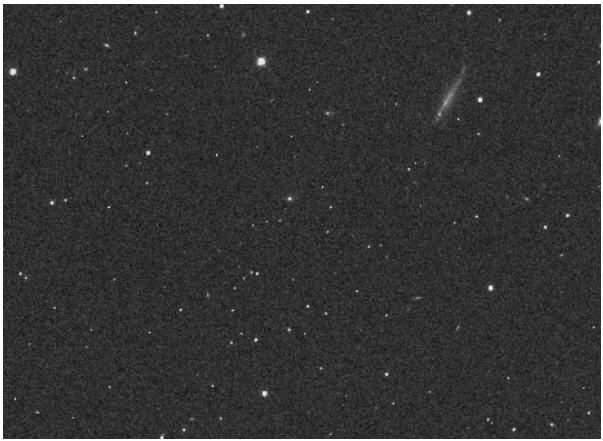}
\caption{An example of the SDSS astronomical image used in comparison test (image info: SDSS telescope 2.5m, exposure time 53.9 sec., RA 194.45516, DEC 0.005886, date: 21/03/99, filter: $g$)}
\end{figure}

\section{Test of interpolation methods}

\subsection{Present methods of bad pixel interpolation}
\label{sec:3.1}
Five interpolation algorithms were compared with respect to pixel brightness estimation in the astronomical images: the median of the surrounding pixels, the nearest neighbor interpolation, the linear interpolation, the cubic spline interpolation and the biharmonic interpolation.
The first and also the simplest method among the presented methods is the nearest neighbors interpolation. The brightness of the pixel in such a case is simply replaced by the brightness of an adjacent pixel. It is to be chosen which of four adjacent pixels to select. Although the selection is arbitrary, it does not influence the overall estimation quality.
The second method uses the median of surrounding pixels. In our tests, for this method, we chose the median of all 8 surrounding pixels' brightness.
The next method is the most popular for bad pixel interpolation in astronomical images - the linear interpolation. It is implemented in widespread astronomical software package IRAF in the {\it fixpix} procedure \citep{mas97}. The interpolation is based on the linear interpolation along columns or rows of the image. In such a case simple mean value of the pixels adjacent to the bad pixel (in a row or in a column) is computed. Again, the decision of choosing row or column interpolation does not change the mean estimation error.
The spline interpolation algorithm uses piecewise polynomials, called splines, to interpolate over the bad pixel along columns or rows. It finds the smoothest curve that passes through data points \citep{ahl67}. In our experiments we used cubic spline interpolation which means that we chose the third order polynomials.
The last method is the biharmonic interpolation. This method has not been used for the bad pixel interpolation yet. It is based on the linear combination of Green functions centered at each data point. Originally, the idea was applied for GEOS-3 and SEASAT altimeter data in 1987 \citep{san87}. 

\subsection{Comparison test}
\label{sec:3.2}
As we mentioned, we used 1000 calibrated astronomical images from SDSS data server. From every single image, two hundreds of 7x7 pixel fragments were extracted. The fragments, containing whole or only parts of astronomical objects (e.g. stars, galaxies), were selected with use of a threshold-based algorithm. It prevented from choosing fragments with the image background. The objects were not centered in each fragment. We did not use any special algorithm for object classification, so both point-like and extended objects were included in our tests. The center pixel of every fragment was chosen as an unknown and its brightness was estimated using previously mentioned interpolation methods. The brightness estimation was compared with the true value to calculate an error. The error absolute and relative value histograms for given interpolation methods are presented in Fig. 3 and 4. Additionally, a mean error ({\it E}) was computed to enable a quick comparison of the methods' effectiveness. The following averaging formula was used:

\begin{equation}
\label{eq1}
E = \sum_{i=1}^N |\Delta_i|
\end{equation}

\noindent
where: $E$  label mean error, $\Delta_i$ --- $i$-th estimation error (absolute or relative) and N --- total number of estimations. We also present point-plots (Fig. 5) to visualize dependency between the estimation error and the brightness of interpolated pixel for all tested methods.

\begin{figure}[H]
\label{fig:2}
\centering
\epsscale{0.50}
\plotone{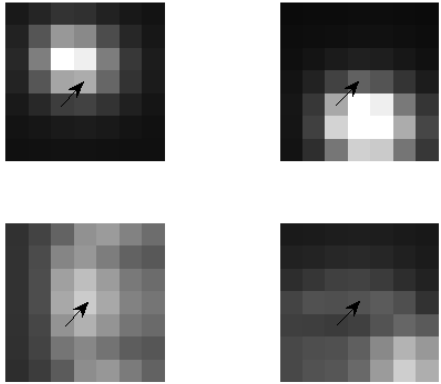}
\caption{Exemplar fragments, 7x7 pixels. Interpolated pixel is pointed by an arrow.}
\end{figure}

There is a clear difference in effectiveness between the used methods. The best results were obtained with the biharmonic interpolation with a mean error equal to 465 [ADU] (ADU stands for Analog-Digital Units and is a measure of signal from a CCD/CMOS pixel) and 0.063 relative, while the typically used method --- linear interpolation --- has an error far greater (872 [ADU] and 0.125 relative). 
The basis and an implementation of the most accurate interpolation are quite simple. According to \citep{san87} the biharmonic interpolation is based on computation of $\alpha_i$ coefficients \eqref{eq2}. For given values of the intermediate points, the system of equations \eqref{eq2} can be easily solved as it is a linear one. 
The biharmonic interpolation as the most accurate method was used as a starting point for further enhancement. It should be also noticed that there is no evidence of applying such an interpolation for astronomical images reduction.

\begin{equation}
\label{eq2}
w(x_j, y_j) = \sum_{i=1}^N \alpha_i \Big( (x_j - x_i)^2 + (y_j - y_i)^2 \Big)   \bigg(2  \ln  \sqrt{(x_j - x_i)^2 + (y_j - y_i)^2} -1 \quad  \bigg)
\end{equation}

\noindent
where: $w(x_i, y_i)$ --- value of the interpolated function for $x_i, y_i$ coordinates, $N$ --- number of intermediate points.

\begin{figure}[H]
\label{fig:3}
\centering
\epsscale{0.75}
\plotone{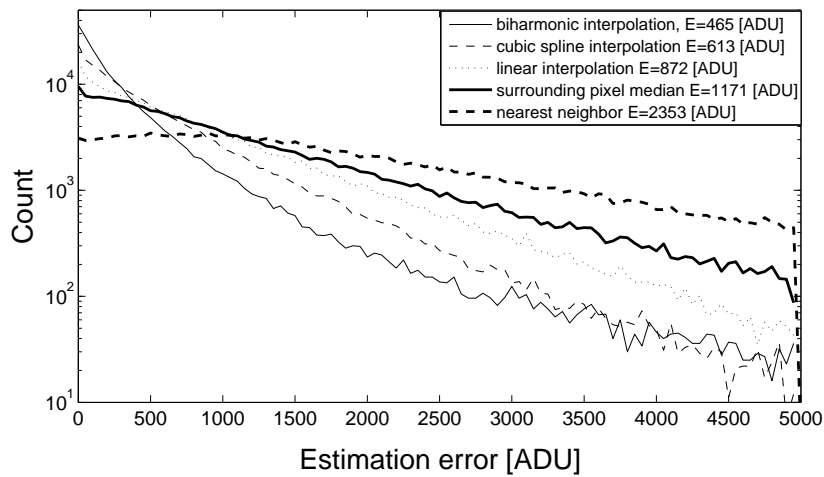}
\caption{Estimation error histogram for the examined interpolation methods. Lines description: solid line -- biharmonic interpolation, dashed line -- cubic spline interpolation, dotted line -- linear interpolation, solid bold line -- surrounding pixel median, dashed bold line -- nearest neighbor.}
\end{figure}

\begin{figure}[H]
\label{fig:4}
\centering
\epsscale{0.75}
\plotone{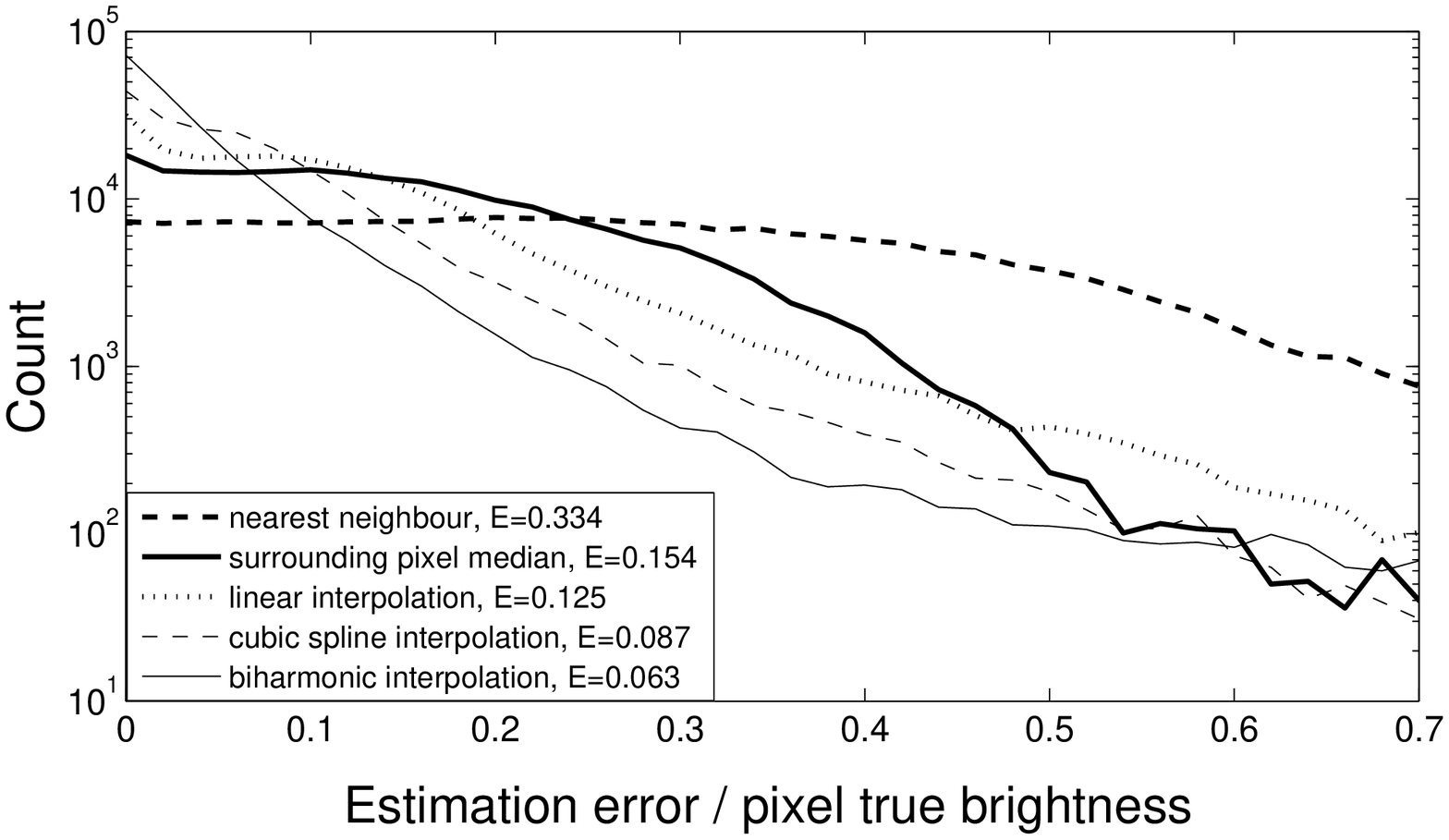}
\caption{Relative estimation error histogram for the examined interpolation methods. Lines description: solid line -- biharmonic interpolation, dashed line -- cubic spline interpolation, dotted line -- linear interpolation, solid bold line -- surrounding pixel median, dashed bold line -- nearest neighbor.}
\end{figure}

\begin{figure}[H]
\label{fig:5}
\centering
\epsscale{0.75}
\plotone{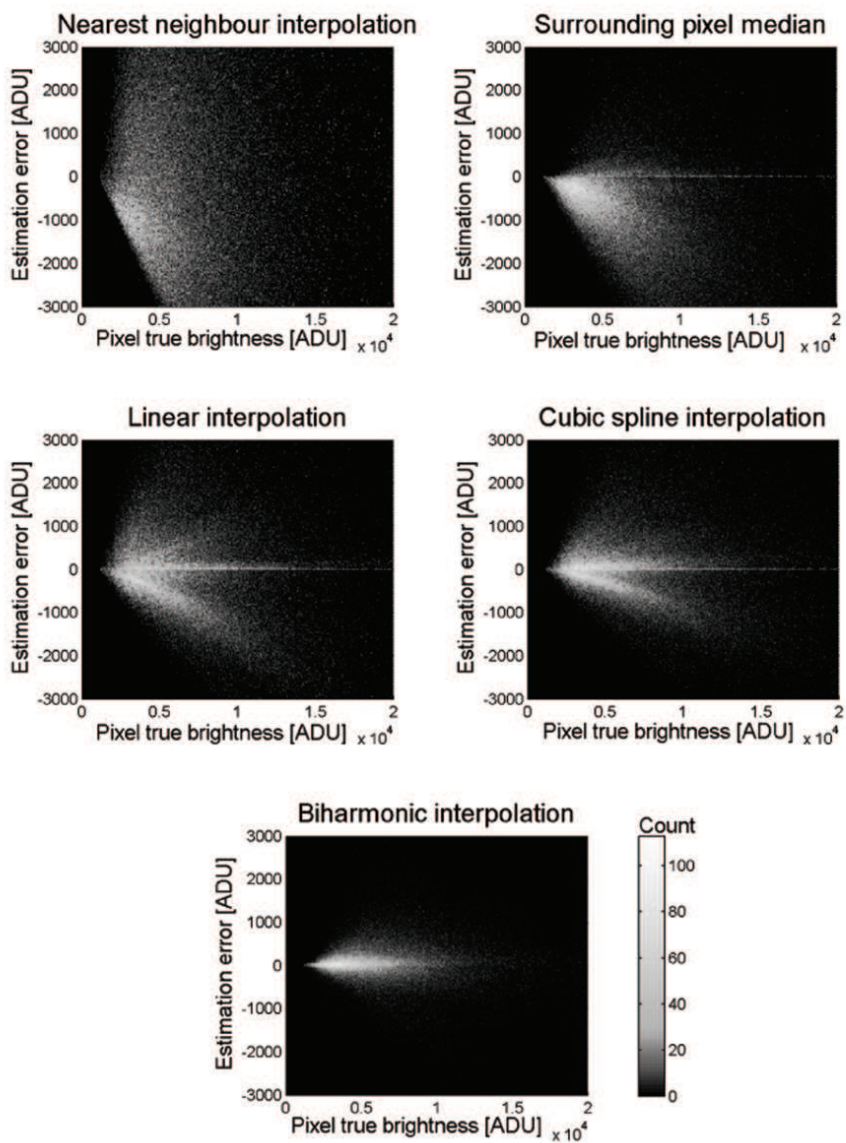}
\caption{Estimation error dependence on interpolated pixel brightness.}
\end{figure}

\section{Modified interpolation}

\subsection{Presentation of the idea}
\label{sec:4.1}
The idea of the modified interpolation is based on the similarity between the objects (both point sources like stars or extended objects like galaxies) encountered in the astronomical pictures. According to \citep{mclen08} a light from a star, as a point source, is affected by the atmosphere (so-called "seeing") and is blurred so that it is registered in the image not as a point but as a point-spread-function (PSF). The PSF can be modeled by the Gaussian surface \citep{pan09}. Usually each star in a high quality telescope system and for more than several seconds exposure shows very similar PSF. However, sometimes PSF can be different due to the seeing variations during the exposure or due to the optical aberrations. Our modified interpolation idea can be applied whether the PSF is stable or not.

We construct the database of 7x7 pixel fragments from the parts of collected images not affected by the bad pixels. Because of the fact that most of the image is usually the background, a special threshold-based selection algorithm is used again to choose as many fragments with the objects as it is possible. Additionally, every fragment is rotated three times (90$^\circ$, 180$^\circ$, 270$^\circ$) to extend the database and to provide for different orientation of extended objects. After the fragments' acquisition, the biharmonic interpolation in every fragment is proceeded over the pixels except for the central pixel. The brightness estimation of the central pixel is compared with the true (already known) value and the estimation error ($e_k$) is computed. The biharmonic interpolation coefficients $(\alpha_{ik})$ and the corresponding estimation error ($e_k$) are stored in the database as a single reference pair. The principle of supporting the interpolation with the database is to compare computed interpolation coefficients with the references from the database and to find the reference which is best suited. The mean square fitting error \eqref{eq3} between the reference and current coefficients is used during the search.

\begin{equation}
\label{eq3}
J_k = \sqrt{ \frac{\sum_{i=1}^N(\alpha_{ik}- \alpha_i^*)^2}{N}     } \quad \xrightarrow{\quad k \quad} \quad min
\end{equation}

\noindent
where: $J_k$ ---  mean square fitting error between $k$-th reference and current coefficients, $ \alpha_{ik}$ --- $i$-th coefficient of $k$-th reference in the database,  $\alpha_i^*$ --- $i$-th coefficient of the interpolation of current fragment and $k$ --- the number of the most suitable reference.

After finding the best reference, the corresponding estimation error is basically subtracted from current estimation.
In sum: the method uses the biharmonic interpolation and is based on a properly created database of known fragments of the image containing astronomical objects. During the interpolation, the database is searched to find the most suitable reference and to apply the corresponding correction to current brightness estimation.
To improve the efficiency of the method, it is desired to find not one but a few (in the experiments the number of 5 fragments was used) of the most suitable references and to subtract a weighted average of the corresponding estimation errors \eqref{eq4}. 

\begin{equation}
\label{eq4}
\overline{e} = \sqrt{ \frac{\sum_{i=1}^5 \frac{1}{J_i} e_i}{ \sum_{i=1}^5 \frac{1}{J_i}}}
\end{equation}

\noindent
where: $\overline{e}$ --- averaged estimation error, $J_i$ ---  mean square fitting error between $i$-th best fitted reference and current coefficients, $e_i$ - estimation error of $i$-th best fitted reference.

Another additional improvement is to normalize every fragment before any interpolation - during the database creation and during a bad pixel correction \eqref{eq5}. The normalization enables to compare fragments of very bright objects with dimmer ones. The bad pixel estimated brightness in such a solution has to be followed by a simple renormalization \eqref{eq6}.

\begin{equation}
\label{eq5}
p_i^{ \prime} = \frac{p_i - min(p)}{max(p) - min(p)} \thinspace  ,
\end{equation}

\begin{equation}
\label{eq6}
p_i = p_i^{ \prime}  \thinspace \big( max(p) - min(p)     \big) + min(p)
\end{equation}

\noindent
where:  $p_i^{ \prime}$ --- normalized pixel brightness; $p_i$ --- real pixel brightness; $min(p)$, $max(p)$ --- minimal and maximal real brightness among the pixels in a fragment.

\subsection{Verification}
\label{sec:4.2}
The database was created using fragments from the first 10 pictures from the previously mentioned SDSS image set. Other images (990 pictures) were used as a verification set. The results are presented for three modified interpolation types (without averaging and normalization, with averaging and with averaging and normalization) and for the original biharmonic interpolation without modification (Fig. 6. and 7).
In another test, it was examined if the modified interpolation can be useful for a single image correction. It means that the database had to be created from the parts of an image that were not affected by faulty pixels. Six randomly chosen images from the SDSS survey were used. 80\% of the 7x7 pixel fragments were used as a database and the remaining 20\% were the test set. The results are resented in Fig. 8. Additionally we compared the effectiveness of the method for the individual frame correction using database constructed from the same frame (minimal database) and from other 20 frames (large database). We analyzed 50 images for this test. The results are presented in Fig. 9. and 10.

The mean error for the test with a large database showed about 50\% decrease after using the modified interpolation idea. According to the histograms (Fig. 6 and 7) the number of small errors rose and the number of large errors decreased as the modified interpolation was used. The proposed enhancements improved also the results of the method. It was shown that both enhancements reduced the mean error by about the same value (20 [ADU]). However, in terms of relative error, the normalization improved the precision only slightly (form 0.044 to 0.042) in comparison to the weighting improvement (from 0.042 to 0.032).

\begin{figure}[H]
\label{fig:6}
\centering
\epsscale{0.75}
\plotone{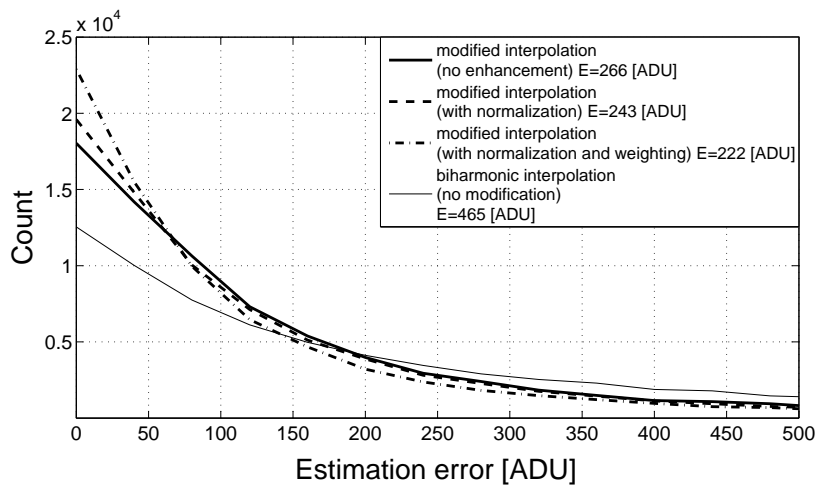}
\caption{Estimation error histogram for modified biharmonic interpolations. Lines description: solid line -- biharmonic interpolation without modification, solid bold line -- modified biharmonic interpolation without enhancements, dashed bold line -- modified biharmonic interpolation with normalization, dash-dot solid line -- modified interpolation with normalization and weighting.}
\end{figure}

\begin{figure}[H]
\label{fig:7}
\centering
\epsscale{0.75}
\plotone{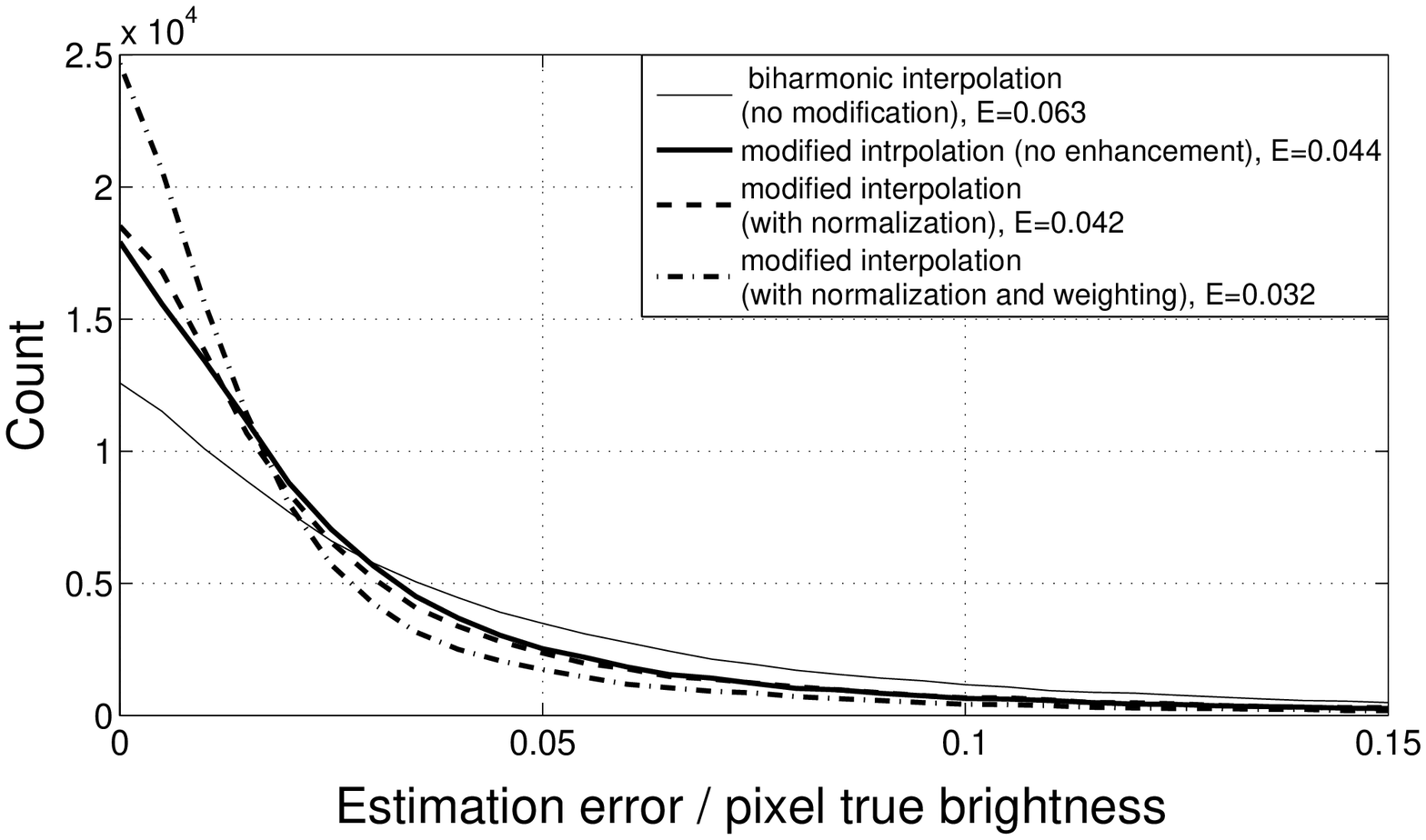}
\caption{Relative estimation error histogram for modified biharmonic interpolations. Lines description: solid line -- biharmonic interpolation without modification, solid bold line -- modified biharmonic interpolation without enhancements, dashed bold line -- modified biharmonic interpolation with normalization, dash-dot solid line -- modified interpolation with normalization and weighting.}
\end{figure}

\begin{figure}[H]
\label{fig:8}
\centering
\epsscale{1.1}
\plotone{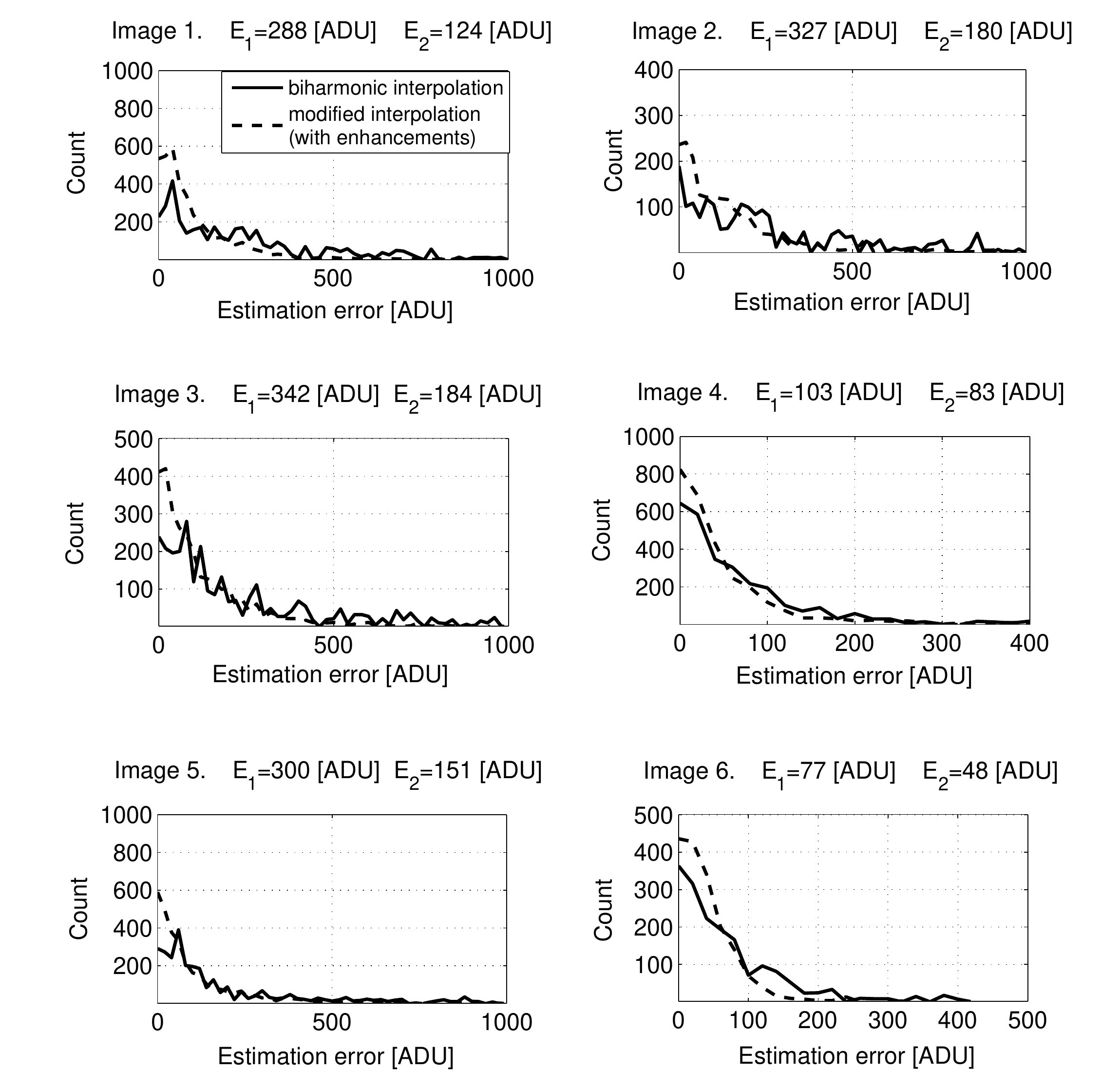}
\caption{Estimation error histogram for modified and not modified interpolation in a single image correction. $E_1$ -- mean estimation error for not modified biharmonic interpolation, $E_2$ -- mean estimation error for modified biharmonic interpolation with enhancements. Lines description: solid line - biharmonic interpolation, dashed line - modified interpolation with enhancements.}
\end{figure}

\begin{figure}[H]
\label{fig:9}
\centering
\epsscale{0.75}
\plotone{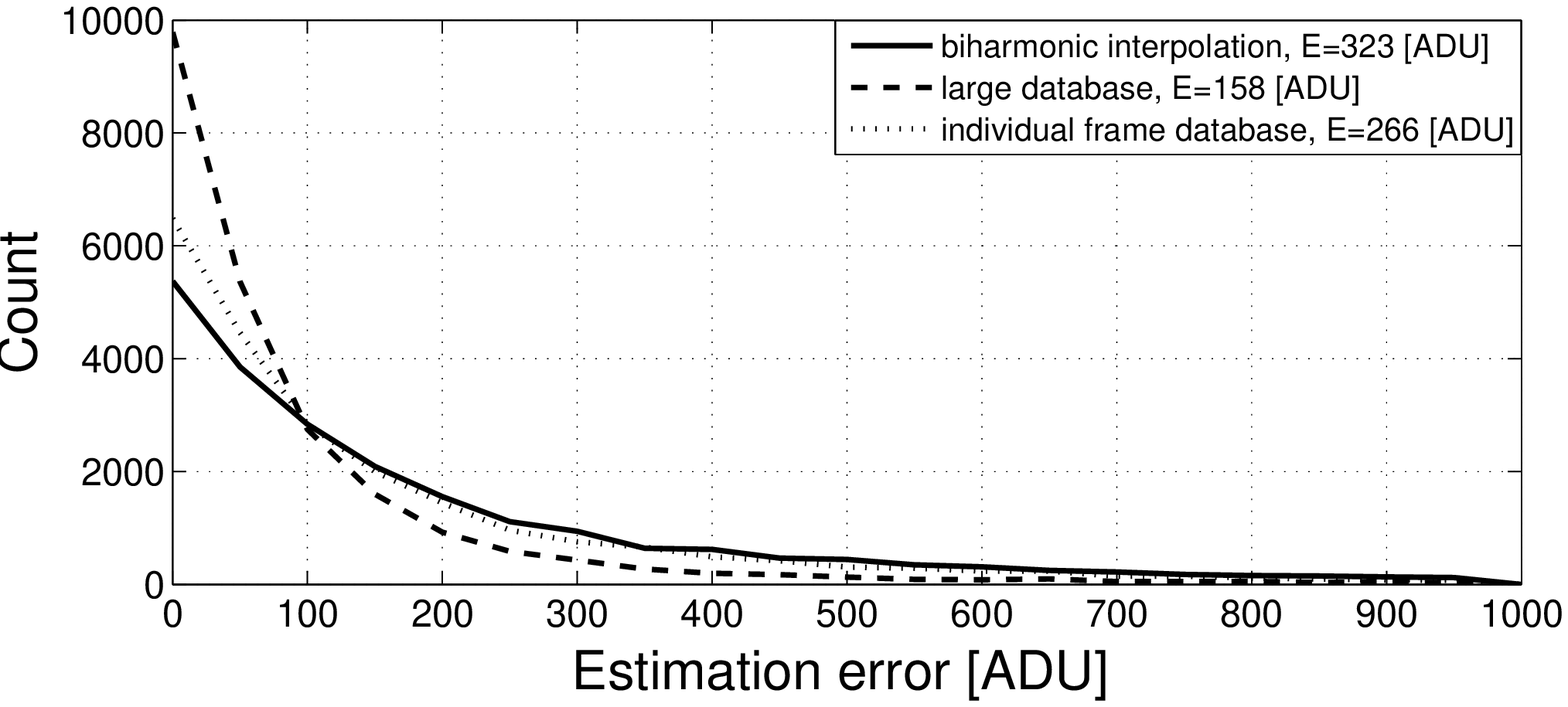}
\caption{Estimation error histogram for individual frame correction. Lines description: solid line -- biharmonic interpolation (no modification), dashed line -- modified biharmonic interpolation with 20 frames database, dotted line --  modified biharmonic interpolation with database constructed from the same frame.}
\end{figure}

\begin{figure}[H]
\label{fig:10}
\centering
\epsscale{0.75}
\plotone{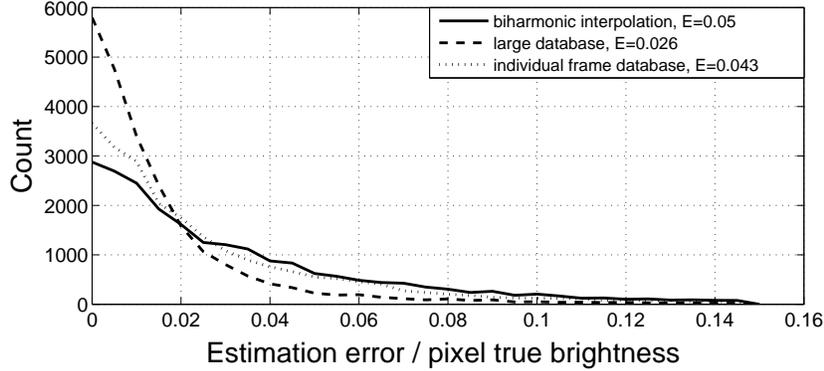}
\caption{Relative estimation error histogram for individual frame correction. Lines description: solid line -- biharmonic interpolation (no modification), dashed line -- modified biharmonic interpolation with 20 frames database, dotted line --  modified biharmonic interpolation with database constructed from the same frame.}
\end{figure}

The modified interpolation seems to work well also as a single image correction, where the database has to be created from fragments of the same image. However, with such a minimal database, the estimation improvement will be strongly dependent on the image and the positions of bad pixels (Fig. 8). For our set of 50 images, the use of larger database improved the estimation error noticeably (Fig. 9 and 10). However it is to be mentioned, that for the images affected by strong PSF variation (e. g. due to the seeing problems), the results would be different.

\section{Conclusions}
\label{sec:5}
In the article a comparison of some interpolation methods of bad-pixels correction in astronomical images was presented. The images from the Sloan Digital Sky Survey were used as an examination set. The biharmonic interpolation as the most accurate method was enhanced with the idea of supporting it with a database of known astronomical image fragments. The test with a large database and a minimal database proved the effectiveness of the method as a pixel's brightness estimator and its superiority over other interpolation methods examined in our tests. Moreover, the biharmonic interpolation has not been used for the astronomical images interpolation yet. With the supporting idea applied, its accuracy was about 4 times higher than for the linear interpolation, which is typically used for the astronomical image calibration. It should be added that the modified interpolation idea is flexible and it could be applied to any current or future interpolation method.

We suggest to consider implementing presented method into data reduction pipelines of large sky surveys. For this kind of application the method should be especially efficient.

\acknowledgments

We want to thank Dr. Agnieszka Pollo (Astronomical Observatory of Jagielonian University) for numerous useful advises. We also thank anonymous referee for useful and constructive comments which allowed to improve the manuscript significantly.

Funding for the SDSS and SDSS-II has been provided by the Alfred P. Sloan Foundation, the Participating Institutions, the National Science Foundation, the U.S. Department of Energy, the National Aeronautics and Space Administration, the Japanese Monbukagakusho, the Max Planck Society, and the Higher Education Funding Council for England.


\begin{thebibliography}{}
\bibitem[Saha(2009)]{saha09} S. K. Saha, "Detectors for the astronomical applications", International Conference on Emerging Trends in Electronic and Photonic Devices \& Systems, Bangalore, India, Dec. 22-24, 2009, pp. 442-445
\bibitem[McLen(2008)]{mclen08} I. S. McLen, "Electronic Imaging in astronomy: Detectors and Instrumentation", Springer-Praxis Publishing, Chichester, UK, 2008
\bibitem[Litwiller(2001)]{lit01} D. Litwiller, "CCD vs CMOS: Facts and Fiction", Photonics Spectra, no. 1, pp. 154-158, 2001
\bibitem[Janesick(2001)]{jan01} J. R. Janesick, "Scientific Charge Coupled Devices", SPIE Press Monograph, Vol. PM83, 2001
\bibitem[Hobson(1978)]{hob78} G. S. Hobson, "Charge-transfer devices", John Wiley and Sons Inc., New York, 1978
\bibitem[Widenhorn(2010)]{wid10} R. Widenhorn, J.C. Dunlap, E. Bodegom, "Exposure Time Dependence of Dark Current in CCD imagers", IEEE Transactions on Electron Devices, vol. 57, no. 3, pp. 581-587, 2010
\bibitem[Dunlap(2012)]{dun12} J. C. Dunlap, M.M. Blouke, E. Bodegom, R. Widenhorn, "Modeling Nonlinear Dark Current Behaviour in CCDs", IEEE Transactions on Electron Devices, vol. 59, no. 4, pp. 1114-1122, 2012
\bibitem[Popowicz(2011a)]{pop11a} A. Popowicz, "CCD dark current analysis", Przeglad Elektrotechniczny (Electrical Review), vol. 87, no. 4, pp. 260-263, 2011
\bibitem[Popowicz(2011b)]{pop11b} A. Popowicz, "A correction algorithm for dark current in CCDs", Przeglad Elektrotechniczny (Electrical Review), vol. 87, no. 11, pp. 318-320, 2011
\bibitem[Hopkinson(2008)]{hop08} G. R. Hopkinson, V. Goiffon, A. Mohammedzadeh, "Random Telegraph Signals in Proton Irradiated CCDs and APS", IEEE Transactions on Nuclear Science, vol. 55, no. 4, pp. 2197-2204, 2008
\bibitem[Hopkinson(1996)]{hop96} G. R. Hopkinson, C.J. Dale, Marshall P.W., "Proton Effects in Charge-Coupled Devices", IEEE Transactions on Nuclear Science, vol. 42 no. 2, pp. 614-627, 1996
\bibitem[Hilbert(2012a)]{hil12a} B. Hilbert, "WFC3/IR Cycle 19 Bad Pixel Table Update", Instrument Science Report WFC3 2012-10", Space Telescope Science Institute, Jun. 08, 2012
\bibitem[Hilbert(2012b)]{hil12b} B. Hilbert, L. Petro, "WFC3/IR Dark Current Stability, Instrument Science Report WFC3 2012-11", Space Telescope Science Institute, Jun. 11, 2012
\bibitem[Fruchter(2002)]{fru02} A. S. Fruchter, R.N. Hook, "Drizzle: A Method for the Linear Reconstruction of Undersampled Images", \pasp, vol. 114, no. 792, pp. 144-152, 2002
\bibitem[Massey(1997)]{mas97} P. Massey, "A User's Guide to CCD Reductions with IRAF", Instruction Manual for IRAF software package, 1997
\bibitem[York(2000)]{yor00} D. G York. et al., "The Sloan Digital Sky Survey: Technical Summary", \apj, vol. 120, pp. 1579-1587, 2000
\bibitem[Szalay(1999)]{szal99} A. S. Szalay, "The Sloan Digital Sky Survey", Computing in Science \& Engineering, vol. 1, no. 2, pp. 54-62, 1999
\bibitem[Ahlberg(1967)]{ahl67} J. H. Ahlberg, E.N. Nilson, J.L. Walsh, "The Theory of Splines and Their Applications" Academic Press, New York, 1967
\bibitem[Sandwell(1987)]{san87} D. T. Sandwell, "Biharmonic Spline Interpolation of GEOS-3 and SEASAT Altimeter Data", Geophysical Research Letters, vol. 14, no. 2, pp. 139-142, 1987
\bibitem[Pan(2009)]{pan09} J. Pan, C. Zhang, "An Efficient Object Detection Method For Large CCD Astronomical Images", Congress on Image and Signal Processing, Jinan, China, Oct. 17-19, 2009, pp.1-4
\end{thebibliography}
\end{document}